\documentclass{aa}
\usepackage{graphicx}
\usepackage{natbib}
\usepackage{rotating}
\usepackage{txfonts}

\bibpunct{(}{)}{;}{a}{}{,}
\def\aap{A\&A}

\def\aj{AJ}
\def\apj{ApJ}
\def\apjs{ApJS}

\def\apjl{ApJ}

\def\mnras{MNRAS}

\def\aapr{A\&ARev}

\def\pasa{PASA}

\def\ssrv{Space Sci. Rev.}

\begin{document}

\title{Hints about the multiplicity of WR\,133 based on multiepoch radio observations}

\author{M. De Becker\inst{1}, N.~L. Isequilla\inst{1,2,3}, P. Benaglia\inst{2,3}}

\offprints{M. De Becker}

\institute{Space sciences, Technologies and Astrophysics Research (STAR) Institute, University of Li\`ege, Quartier Agora, 19c, All\'ee du 6 Ao\^ut, B5c, B-4000 Sart Tilman, Belgium
\and
Instituto Argentino de Radioastronom\'{\i}a (CONICET;CICPBA), C.C. No 5, 1894, Villa Elisa, Argentina
\and
Facultad de Ciencias Astron\'{o}micas y Geof\'{\i}sicas, UNLP, Paseo del Bosque s/n, 1900, La Plata, Argentina
\\
}

\date{Received ; accepted }

\abstract
{Several tens of massive binary systems display indirect, or even strong evidence for non-thermal radio emission, hence their particle accelerator status. These objects are referred to as particle-accelerating colliding-wind binaries (PACWBs). WR\,133 is one of the shortest period Wolf-Rayet + O systems in this category, and is therefore critical to characterize the boundaries of the parameter space adequate for particle acceleration in massive binaries. Our methodology consists in analyzing JVLA observations of WR\,133 at different epochs to search for compelling evidence for a phase-locked variation attributable to synchrotron emission produced in the colliding-wind region. New data obtained during two orbits reveal a steady and thermal emission spectrum, in apparent contradiction with the previous detection of non-thermal emission. The thermal nature of the radio spectrum along the 112.4-d orbit is supported by the strong free-free absorption by the dense stellar winds, and shows that the simple binary scenario cannot explain the non-thermal emission reported previously. Alternatively, a triple system scenario with a wide, outer orbit would fit with the observational facts reported previously and in this paper, albeit no hint for the existence of a third component exists to date. The epoch-dependent nature of the identification of synchrotron radio emission in WR\,133 emphasizes the issue of observational biases in the identification of PACWBs, that undoubtedly affect the present census of PACWB among colliding-wind binaries.}


\authorrunning{De Becker et al.}
\titlerunning{Radio emission from WR~133}

\maketitle

\section{Introduction}\label{intro}
The radio investigation of massive stars, that is, OB-type and Wolf-Rayet (WR) objects, revealed both thermal emission and non-thermal emission spectra \citep[see e.g.,][]{ABC,ABCT, montes2009}. The thermal emission is free-free radiation from the optically thick stellar winds as described by \citet{PF} and \citet{WB}, characterized by a spectral index $\alpha \sim 0.6$ (for a frequency dependence of the flux density defined as $S_\nu \propto \nu^\alpha$). The non-thermal emission is synchrotron radiation produced by relativistic electrons in the presence of a magnetic field \citep{Wh}, characterized by a negative spectral index. In this context, flatter spectra suggest a combination of unresolved thermal and non-thermal sources, the so-called composite spectra. The particle acceleration process responsible for the presence of relativistic electrons is believed to be diffusive shock acceleration \citep{bella,drury1983}, in the presence of the strong shocks produced by the collision of stellar winds in binary system \citep{EU,debeckerreview}. The subset of colliding-wind binaries known to accelerate particles is now referred to as the class of particle-accelerating colliding-wind binaries (PACWBs). The importance of binarity (or even higher multiplicity) for particle acceleration is emphasized in the catalogue of PACWB published by \citet{catapacwb}, where almost all systems show compelling evidence for multiplicity. To date, the detection of synchrotron radiation is the most efficient tracer of particle acceleration in massive binaries.

One striking fact, revealed by the inspection of the catalogue of PACWBs, is the spread of stellar and orbital parameters, suggesting the usual parameter space populated by a large fraction of massive binaries could be adequate for particle acceleration \citep{DBRP}. It is instructive to investigate the apparent limits of this parameter space, in order to achieve a better view of its extension. In this context, the lower period systems carry a significant information. First of all, the synchrotron emission is very likely to be significantly absorbed by the optically thick wind material in small size systems, preventing their identification as particle accelerators. Second, the efficient cooling of relativistic electrons by inverse Compton scattering constitutes a strong inhibition for electron acceleration, and this effect is enhanced in short period systems where the colliding wind region is closer to the stellar photospheres \citep[see][for a discussion]{DBRP}. A better view of the circumstances likely to allow for particle acceleration in shorter period systems is strongly needed to improve our understanding of the non-thermal physics in massive stars. 

Among O-type stars, the shortest period system with known synchrotron radiation coming from a well-identified colliding-wind region is Cyg\,OB2\,\#8A, with a period of about 22 days \citep{Let8a,Blomme8apaper}. Among WR systems, the shortest period referenced in the catalogue of PACWBs is WR~11, with a period of about 80 days \citep{gam2velorbsol}. However, its non-thermal radio emitter status has recently been questioned by the recent radio investigation by \citet{benagliawr11}, though it still may be a particle accelerator provided the Fermi $\gamma$-ray source reported by \citet{pshirkov2016} is physically related to WR~11. 

The present study is dedicated to the next system in order of increasing period in the catalogue: WR~133. The paper is organized as follows. In Sect.\,\ref{target}, we briefly describe the target and the reasons justifying a deeper investigation of its particle accelerator status. Section\,\ref{radio} details the processing and analysis of the data obtained with the Karl G. Jansky Very Large Array (JVLA). A general discussion is given in Sect.\,\ref{disc} and the conclusions are presented in Sect.\,\ref{concl}.

\section{The massive binary WR~133}\label{target}

WR~133 (\object{HD 190918}) consists of a WN5 star with an O9I companion. \citet{UH1994wr133} reported on an eccentric (e = 0.39) orbit with a period of 112.4\,d, which is much shorter than most orbital periods identified for WR systems in the catalogue of PACWBs. The minimum masses and the size of the semi-major axis for both components of the system can easily be estimated on the basis of the results published by \citet{UH1994wr133} and are given in Table\,\ref{param}. On the basis of the calibration of stellar parameters for O9I stars given by \citet{martins}, the expected mass of the O-type component should be close to 30\,M$_\odot$. Comparing this quantity to the minimum mass value given in Table\,\ref{param} one estimates the inclination angle ($i$) to be on the order of 18$^\circ$. This angle allows to estimate the absolute semi-major axis starting from the projected value published by \citet{UH1994wr133}. In the absence of astrometric measurement of the orbit, this rough approach allows to achieve a reasonable view of its actual size.

\begin{table*}
\caption{Adopted parameters for WR~133. \label{param}}
\begin{center}
\begin{tabular}{l c c c c c c c c}
\hline
 & \multicolumn{3}{c}{System parameters} & & \multicolumn{4}{c}{Wind parameters}\\
 \cline{2-4}\cline{6-9}
 \vspace*{-0.2cm}\\
 & $M\,\sin^3 i$ & $M$ & $a$ & & T & ${\dot M}$ & $V_\infty$ & $P_{kin}$ \\
 & (M$_\odot$) & (M$_\odot$) & (R$_\odot$) & & (K) & (M$_\odot$\,yr$^{-1}$) & (km\,s$^{-1}$) & (erg\,s$^{-1}$) \\
\hline
\vspace*{-0.2cm}\\
O9I & 0.825$^{~(1)}$ & 30$^{~(2)}$ & 34.6 & & 15000$^{~(3)}$ & 4.1\,$\times$\,10$^{-7}$$^{~(4)}$ & 3000$^{~(3)}$ & 1.2\,$\times$\,10$^{36}$ \\
WN5 & 0.4$^{~(1)}$ & 16 & 70.3 & & 30000$^{~(3)}$ & 6.3\,$\times$\,10$^{-6}$$^{~(5)}$ & 1500$^{~(5)}$ & 4.5\,$\times$\,10$^{36}$ \\
\vspace*{-0.2cm}\\
\hline
\end{tabular}
\tablefoot{References. (1) \citet{UH1994wr133}, (2) \citet{martins}, (3) based on $T_{eff}$ taken from \citet{martins} or \citet{crowtheraraa}, (4) \citet{muijres}, (5) \citet{crowtheraraa}.}
\end{center}
\end{table*}

The mass loss rate and the terminal velocity for the O9I star were retrieved from \citet{muijres}. For the WR component, we use the typical parameters for a WN5 star given by \citet{crowtheraraa}. Considering the wind parameters given in Table\,\ref{param}, it is clear that the wind kinetic power of the WN wind dominates the wind collision that is located closer to the O-star. The wind temperature (T) is assumed to be 0.5\,$\times$\,$T_{eff}$ \citep{drew1989}, and will be used in Sect.\,\ref{discbin}.\\

\begin{figure}[ht]
\begin{center}
\includegraphics[width=85mm]{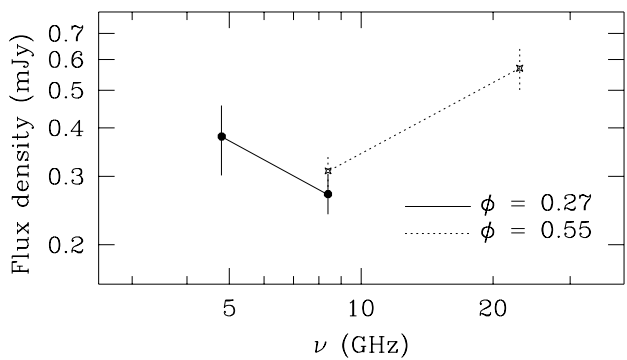}
\caption{Radio measurements published by \citet{montes2009}. Measurements at both epochs are overplotted. \label{radiomontes}}
\end{center}
\end{figure}

Previous radio measurements presented by \citet{montes2009} revealed a likely composite spectrum, that is, made of thermal and non-thermal emission without any strong and clear signature for a radio emission dominated by synchrotron radiation. This is not surprising as the small size of the system barely allows synchrotron photons to escape efficiently from the combined optically thick winds in the system, especially considering the WN component. The flux densities reported by \citet{montes2009} are 0.31\,$\pm$\,0.03\,mJy and 0.57\,$\pm$\,0.07\,mJy at 8.4\,GHz and 23\,GHz, respectively, with an upper limit of 0.41\,mJy at 4.8\,GHz. The spectral index is equal to 0.6, that is, the expected value for a purely thermal emission from a massive star wind \citep{PF,WB}. According to the ephemeris published by \citet{UH1994wr133}, the observation date (6 May 2007) corresponds to orbital phase $\phi$ = 0.55, shortly after apastron passage. However, measurements at 4.8\,GHz and 8.4\,GHz (0.38\,$\pm$\,0.08\,mJy and 0.27\,$\pm$\,0.03\,mJy, respectively) performed on 31 May 1993 ($\phi$ = 0.27) suggest a spectral index of --0.61\,$\pm$\,0.43, typical of many non-thermal sources. Though the error bar on the spectral index is large, the confidence interval still indicates at least a composite spectrum, and rejects a pure thermal emission. Measurements by \citet{montes2009} are represented in Fig.\,\ref{radiomontes} as a function of frequency. The different spectral nature roughly below and above 8.4\,GHz could be interpreted in terms of composite spectrum, with the lower frequencies revealing some synchrotron emission, and the higher frequency domain dominated by thermal emission mainly from the WN stellar wind. One also has to caution that both epochs correspond to different orbital phases, and the difference may also be attributed to an orbital phase-locked variation. 

On the other hand, the time interval of more than 20 years between both observations raises the question of a potential longer term variability of the radio emission, independently of the expected variation along the 112 days orbit. The only point of comparison is the measurement at 8.4 GHz obtained at both epochs. The flux densities are equivalent within error bars, but they correspond to different orbital phases. One cannot therefore conclude on the long term behavior of the radio spectrum of WR\,133 on the basis of these previous observations only. Moreover, one has also to envisage the possibility that the uncertainties on the ephemeris published by \citet{UH1994wr133} propagate significantly in the determination of the orbital phases of the observations. More than 25 years elapsed since the reference time of the ephemeris at the time of our new JVLA observations, corresponding to more than 80 orbits of the system. One cannot thus reject the idea that the orbital phases are not correct. Without any more accurate determination of the ephemeris, one can only speculate on the potential impact of this error propagation. However, the spread of our new time series (see Fig.\,\ref{orbit}) is enough to sample the 112.4-d orbit, whatever the impact of this ephemeris error propagation. A revision of the orbital solution based on additional spectroscopic observations to be used altogether with the \citet{UH1994wr133} is thus necessary. In particular, a significantly much longer time basis of radial velocity measurements would allow to reduce significantly the error on the period, and reduce the uncertainties on the orbital phases of the JVLA observations as well.\\

Our motivation is to monitor WR\,133 to connect the previous indication of non-thermal radio emission to its orbital phase-dependent behavior, as expected for an eccentric binary system. This would allow in principle to clarify the role of free-free absorption as a function of the stellar separation along the orbit, which is anticipated to be quite strong in this system (strong stellar winds, rather short period). A phase-locked variation would constitute a strong support for the colliding-wind origin of the synchrotron radiation in WR\,133. Any deviation from this anticipated behavior would trigger new questions for our understanding of this system, and by extension open prospects for future investigations.

\section{Radio observations of WR~133}\label{radio}

At our request, WR~133 was observed with the JVLA in configuration A at C and X bands, at various epochs (Obs. Id 14A-074 and 15A-023; see Table\,\ref{obs}), respectively in 2014 and 2015. The orbital phases of the observations determined on the basis of the ephemeris of \citet{UH1994wr133} are given in Table\,\ref{obs}.

The observations of 2015 were made at the frequency bands centered at 5.5 GHz (band 1) and 9 GHz (band 2), along a 2 GHz bandwidth, while 2014 observations were made in band 1 only. The data were reduced with the Common Astronomy Software Applications package (CASA, \citealt{CASA}). The source 3C295 was used as primary calibrator, with fluxes of 5.582 Jy at 5.488 GHz (C band) and 3.077 at 8.988 GHz (X band) \citep{PerleyButler2013}. J2015\,+3710 served as the phase calibrator. At each frequency band, the bandwidth comprised 16 spectral windows of 64 channels each. The time on source at each band and epoch was 16.5 minutes in 2015, and 5.2 minutes in 2014 in band 1. The theoretical image noise was 15 microJy/beam in 2014 and 8 microJy/beam in 2015. 
 
\begin{figure}[ht]
\begin{center}
\includegraphics[width=9cm]{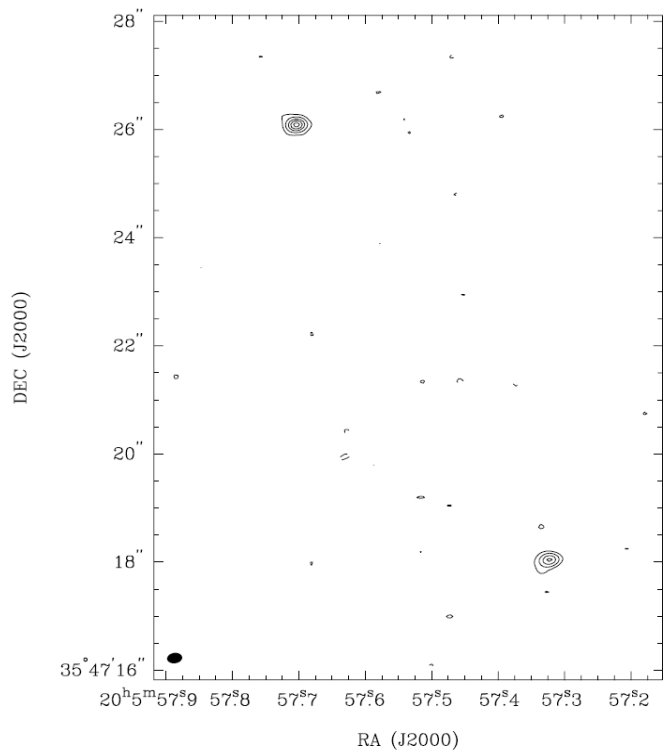}
\caption{Radio image showing the two point sources detected in the JVLA data at 9\,GHz, in July 2015, WR\,133 (bottom) and the NE source (top). The black ellipse at the bottom left corner illustrates the beam size. The contour levels are —0.025, 0.025, 0.08, 0.16, 0.24 and 0.32 mJy beam$^{-1}$.}
\label{imagetot}
\end{center}
\end{figure} 
 
Bad data were flagged, and the images were built using robust weightings\footnote{Models for 3C295 were kindly provided by D. Medlin and R. Perley.}. 
The final images show an rms between 10 to 20 microJy/beam. WR\,133 is detected in all images. 
The synthesized beams and the rms attained of each image are given in Table\,\ref{obs}.
The July image obtained at 9 GHz is displayed in Fig.\,\ref{imagetot}. Two point sources are detected in the field, our target (WR\,133) and an additional source northeast from WR\,133. Since the synthesized beam is larger than the true dimensions of the sources in both bands, the size of the imaged sources is that of the synthesized beam. The flux density of WR 133 was estimated by fitting Gaussians, and the measured values are quoted in Table\,\ref{obs}. For data analysis we also used the Miriad package \citep{MIRIAD} and the karma software \citep{KARMA}. The spectral index for the 2014 and 2015 observations was determined using the equation $\alpha = \log\,(S_{\nu,1}/S_{\nu,2})/\log\,(\nu_1/\nu_2)$, and the error bar was estimated on the basis of the following equation:
\begin{equation}
\Delta\alpha = \sqrt{\bigg[\frac{1}{\ln(\nu_1/\nu_2)}\bigg]^2\,\Bigg[\bigg(\frac{\Delta\,S_{\nu,1}}{S_{\nu,1}}\bigg)^2 + \bigg(\frac{\Delta\,S_{\nu,2}}{S_{\nu,2}}\bigg)^2\Bigg]}.
\end{equation}
\noindent The spectral index measurements we determined here are thus inter-band values. Considering the signal-to-noise ratio on the data, any measurement of in-band spectral index would lack the required accuracy to carry any relevant information. In addition, no specific in-band trend different from the inter-band one is expected.

We also compared our new data to observations obtained earlier by \citet{montes2009}, for which the VLA interferometer bandwidth was 50 MHz. To do so, we selected the spectral window (spw) that fell near or within the same central frequencies observed by \citet{montes2009}, which were 4.8 and 8.4 GHz. For the 5.5 GHz frequency data, we chose spw = 1 - 3 (4616–5000 MHz, tagged subb1) near 4.8 GHz. For the 9 GHz band, we chose spw = 2 - 4 (8244–8628 MHz, tagged subb2) near 8.4 GHz. The images were obtained using CASA. Figure\,\ref{imagetot} shows the image for the Jul 2015 epoch data, that showed lower noise compared to the Sep 2015 ones. The flux densities of WR\,133 were estimated using the MIRIAD package. The flux of WR\,133, for Jul 2015, at subb2 is $0.27\pm0.03$ mJy, in accordance with the corresponding values reported by \citet{montes2009}. The measurement at subb1 band yielded a target flux of $0.21\pm0.02$ mJy. Although this result agrees with the flux upper limit derived from 2007 data presented by \citet{montes2009} (and also with the upper limit presented by \citealt{ABCT}), it is remarkably lower than the flux reported from the 1993 data also by \citet{montes2009}. 

We also note that we reprocessed the 1993 data in order to achieve the full confirmation of the non-thermal emitter status of WR\,133 at that epoch, as this constitutes a pivotal element in our understanding and interpretation of the radio properties of this system. Our analysis led to flux densities of 0.38 and 0.28\,mJy, respectively at 4.8 and 8.4\,GHz. These values are in perfect agreement with the results published by \citet{montes2009}, and plotted in Fig.\,\ref{radiomontes}. The important information that will be considered in this study is the indication for non-thermal radio emission previously reported by \citet{montes2009} for WR\,133, on the basis of the 1993 measurements, which appears as a highly valid and consolidated fact.

\begin{table*}
\caption{Journal of observations and flux density measurements of WR\,133. \label{obs}}
\begin{center}
\begin{tabular}{l c c c c c c c}
\hline
Epoch & Julian date & Orb.phase & Band $\nu$ & Synth.beam & rms & S$_\nu$ & $\alpha$\\
 & & ($\phi$) & (GHz) & (",",deg) & (mJy/beam) &  (mJy) & \\
\hline
\vspace*{-0.2cm}\\
27 Feb 2014 & 2456716.5 & 0.70 & 5.5 & $0.63\times 0.33, -64.4$ & 0.02 & 0.21\,$\pm$\,0.02 & --\\
28 Feb 2014 & 2456717.5 & 0.71 & 5.5 & $0.65\times 0.31, -66.2$ & 0.02 & 0.23\,$\pm$\,0.02 & --\\
13 Apr 2014 & 2456761.5 & 0.10 & 5.5 & $0.67\times 0.33, -65.3$ & 0.02 & 0.21\,$\pm$\,0.02 & --\\
23 May 2014 & 2456801.5 & 0.46 & 5.5 & $0.63\times 0.37, -63.7$ & 0.02 & 0.21\,$\pm$\,0.03 & --\\
24 Jul 2015 & 2457227.6 & 0.25 & 5.5 & $0.52\times 0.31, -79.1$ & 0.01 & 0.19\,$\pm$\,0.02 &  \\
 &  &  & 9.0 & $0.28\times 0.19, -82.2$ & 0.01 & 0.27\,$\pm$\,0.02 & $+0.71\,\pm\,0.26$ \\
20 Sep 2015 & 2457285.6 & 0.77 & 5.5 & $0.34\times 0.33, +41.8$ & 0.01 & 0.21\,$\pm$\,0.02 & \\
 &  &  & 9.0 & $0.20\times 0.19, -51.9$ & 0.01 & 0.27\,$\pm$\,0.02 & $+0.51\,\pm\,0.30$\\
\hline
\end{tabular}
\end{center}
\end{table*}

\section{Discussion}\label{disc}
\subsection{The orbit of WR133}\label{orbitsect}

\begin{figure}[ht]
\begin{center}
\includegraphics[width=80mm]{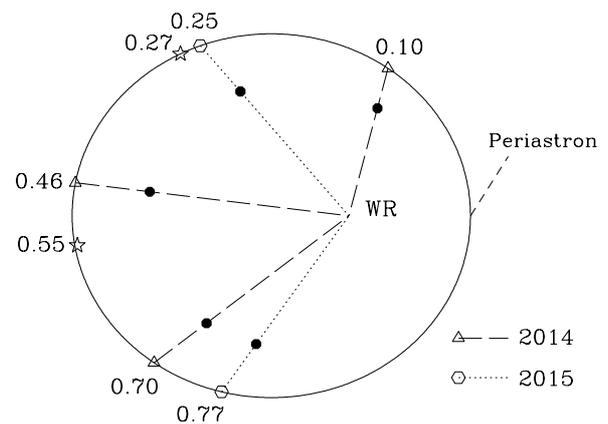}
\caption{Illustration of the orbit of WR~133. The open hexagons and triangles represent the position of the O-star, respectively during two different orbits. The filled symbol stands for the expected position of the wind collision region. The long-dashed and dotted lines represent the stellar separation, in 2014 and 2015 respectively. The orbital phases of the \citet{montes2009} observations are illustrated by the two open star symbols. \label{orbit}}
\end{center}
\end{figure}

The radio measurements should be adequately interpreted in the context of the orbital configuration. Figure\,\ref{orbit} represents the orbit of WR~133 based on the parameters given by \citet{UH1994wr133} and shows the relative position of the stars at the phases of our JVLA observations. For the sake of completeness, we also show the stellar positions at the epoch of the \citet{montes2009} observations. One should clarify that the reference time for periastron passage determined by \citet{UH1994wr133} is located in 1987. As a result the 1993 measurements (where the non-thermal emission is measured, at $\phi = 0.27$) is affected by the lowest uncertainty on the orbital phase determination, while our most recent measurements can in principle be more affected. Our JVLA measurements were indeed obtained about 25 years after the reference time, translating into about 80 full elapsed orbital periods (whose error bar is about 0.2\,d according to \citealt{UH1994wr133}). In the worst (an unrealistic) situation it may lead to an uncertainty on the orbital phase determination of several percent. Anyway, as stated in Sect.\,\ref{target}, these uncertainties do not question the nature of our sampling of the 112.4-d orbit with our new measurements. On the basis of the values given in Table\,\ref{param}, the absolute median separation is $a = a_\mathrm{O}$ + $a_\mathrm{WR}$ = 104.9\,R$_\odot$, with extreme values $a\,(1 - e)$ and $a\,(1 + e)$ respectively at periastron and apastron.

As a first approach, let us make use of the wind parameters quoted in Table\,\ref{param}. We note that the uncertainties on the terminal velocities should be of the order of 5--10\,$\%$ at most, if one considers an uncertainty on the stellar classification of one subtype or one luminosity class (according to the references quoted in the Table). The same is true for the O-star mass loss rate. For the WN star, the uncertainty on the mass loss rate may reach a few tens of $\%$. Such uncertainties are inherent to usual stellar parameters of stellar winds of massive stars, with no significant impact on our conclusions. These parameters allow in particular to compute the wind momentum rate ratio ($\eta$) according to the following relation: $\eta = \big({{\dot M}\,V_\infty}\big)_{\mathrm{O}}/\big({{\dot M}\,V_\infty}\big)_{\mathrm{WR}}$,
 leading to the value $\eta = 0.13$. The position of the wind collision region (WCR) measured from the center of mass of the two stars can then be calculated: $r_\mathrm{WR} = a\,r_\mathrm{rel}/(1 + \sqrt{\eta})$ and $r_\mathrm{O}  = a\,r_\mathrm{rel}\,\sqrt{\eta}/(1 + \sqrt{\eta})$, where $r_\mathrm{rel} = (1-e)/(1+e\,\cos\upsilon)$ and $\upsilon$ being the true anomaly. These positions are also represented in the plane of the orbit in Fig.\,\ref{orbit}.\\

The size of the system is rather small, that is, less than an Astronomical Unit, placing the wind collision region in the wind acceleration zone of the O-type star. A kinematic adjustment should thus be considered, taking into account the `actual' pre-shock velocity instead of assuming terminal values, especially close to periastron passage. Such a kinematic adjustment leads the WCR to get closer to the O-star. Unstable solutions are expected as the distance between the O-star and the WCR calculated following this simple kinematic approach can even become smaller than the expected stellar radius, that is, about 23\,R$_\odot$ \citep{martins}. Beside, the estimated position of the WCR is sensitive to the adopted wind parameters, in particular those of the WR component which are quite uncertain. Even lowering its terminal velocity down to 1000\,km\,s$^{-1}$ does not solve this issue. On the other hand, a higher mass loss rate for the WR could not be completely rejected (see for instance WN wind parameters given by \citealt{HK1998WN}), widening further the orbital phase interval around periastron where the WR wind seems at first sight to crash onto the companion's surface. One should however note that this approach does not take into account radiative effects such as radiative inhibition \citep{radinhib} and sudden radiative breaking \citep{radbraking}, likely to prevent the WR wind to crash onto the O star surface. In addition, the simple description provided here does not take into account the orbital asymmetry due to the orbital motion that distorts the WCR. The full radiative and hydrodynamic treatment of these issues is beyond the scope of the present work. The values quoted above for the position of the wind-wind interaction region should thus be considered as reasonable guess values adequate for the present qualitative discussion.

\subsection{Radio emission from WR\,133}\label{radiodisc}
The flux densities quoted in Table\,\ref{obs} suggest a constant radio emission along the 112.4-d orbit. The spectrum seems to be purely thermal. These results are at odd with those published by \citet{montes2009} from VLA data, presenting the signature of a non-thermal emission component (see Sect.\,\ref{target}).
We note that the observations presented here were accomplished with the (upgraded) Jansky Very Large Array. The flux models and scale of the primary calibrator used in the present investigation have been widely extended in frequency and rebuilt \citep{PerleyButler2013}. In the light of these changes, a meaningful comparison between our values and the ones in the literature is doubtful. The fluxes derived prior to 2010 corresponded to a bandwidth of 50 MHz, while the ones presented here were taken along a 2-GHz band. 
In addition, one also has to take into account the expected strong free-free absorption in that system, especially considering the dense WR wind likely to attenuate dramatically any radio emission produced close to the wind-wind interaction region. For these reasons, the interpretation of the radio behavior is not straightforward and it is worth considering two distinct potential interpretation frameworks, in other words the binary system and the triple system scenarios. These scenarios are critically discussed below, and summarized in Table\,\ref{comparison}. 


\subsubsection{The binary system scenario}\label{discbin}
In this scenario, the thermal and non-thermal emissions reported by \citet{montes2009} and by us are produced in the WCR in the 112.4-d orbit. The rather small size of the orbit constitutes a severe issue considering the significant impact of free-free absorption by the stellar winds material. As a rough estimate of the impact of the free-free absorption by the stellar winds in the system, one can use the formalism developed for instance by \citet{WB} and \citet{leietal} to estimate the typical radius of the radio photosphere of a given stellar wind. The radius of the radio photosphere is defined for an optical depth equal to one, and can be computed following the approach already applied in previous studies \citep[e.g.,][]{DeB168112,mahyhd150136}. 

\begin{figure}[ht]
\begin{center}
\includegraphics[width=85mm]{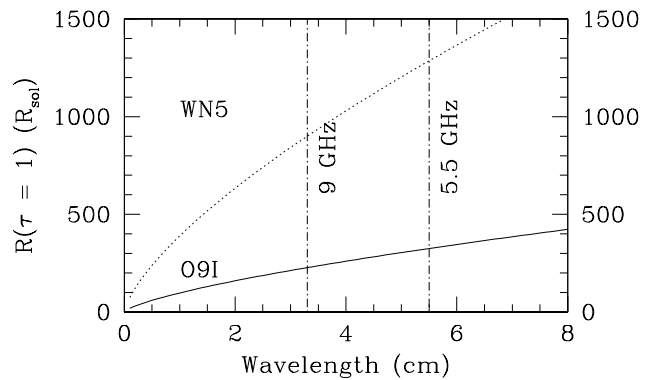}
\caption{Radio photosphere radius as a function of wavelength for both star winds in the WR\,133 system. Vertical lines are located at wavelengths corresponding to our two observation bands. \label{photwr133}}
\end{center}
\end{figure}

Using the parameters given in Table\,\ref{param}, we derived curves for the radio photosphere radius as a function of wavelength for both stars in the system (see Fig.\,\ref{photwr133}). The radio photosphere radii at 5.5 and 9 GHz are significantly larger than the size of the orbit, confirming the a priori idea that a substantial amount of the radio photons produced at the position of the WCR should be absorbed. This is especially true for the WN5 curve, lending support to the idea that the WR stellar wind dominates the free-free absorption in this system. We note that within the uncertainties on the stellar wind parameters, in particular for the WN star (see Sect.\,\ref{orbitsect}), predictions on the radio photosphere radii would still be much too large to admit a potential escape of a measurable fraction of the putative synchrotron emission from the colliding-wind region in the 112.4-d orbit. For instance, lowering the mass loss rate of the WN star by a factor 2 would lead to a drop of the WN curve plotted in Fig.\,\ref{photwr133} by a factor of about 1.6 only. A much more severe change of the mass loss rate would be required, and as stated previously this is very unlikely that such a change in mass lass loss rate occurred between 1993 and the epoch of our new observations. One should keep in mind that the optical depth considered here is radial (i.e., centered on the star), and not measured along the line of sight toward the synchrotron emission region. The measurement of the line-of-sight optical depth would require a much more sophisticated treatment based on detailed hydrodynamic and radiative codes, out of the scope of the present study. One has however to mention the case of another system where the free-free absorption is a priori strong enough to prevent the detection of synchrotron emission: Cyg\,OB2\,\#8A. The latter system presents a clear non-thermal radio signature phase-locked with the 22-days orbital period \citep{Let8a,Blomme8apaper}, providing evidence that the radio photosphere extension beyond the orbit size is certainly a pessimistic criterion. However, the comparison of the numbers plotted in Fig.\,\ref{photwr133} to the size of the orbit (at most $\sim$ 105 R$_\odot$) clearly emphasizes the crucial impact of free-free absorption in this system, which is undoubtedly the dominant turn-over process preventing any obvious identification of synchrotron radiation from WR\,133. In addition, the free-free absorption of any putative synchrotron emission component will be dependent on the orientation of the system. We are lacking the adequate information to achieve a detailed three-dimensional view of the orbit (i.e., the longitude of ascending node), but we estimate the inclination of the system to be of the order of 18$^\circ$ (see Sect.\,\ref{target}). It is not sure that such an inclination is low enough to favor the potential detection of a fraction of the synchrotron emission.

The most obvious source of variation in an eccentric system is the changing stellar separation along the orbit. One could a priori expect the detection of a non-thermal emission close to apastron, where a larger fraction of the synchrotron emission region may emerge from the opaque WR wind. However, even at orbital phase $\phi$ = 0.46 (i.e., the closest to apastron in our time series), we do not find any indication of non-thermal emission. Our sampling of the radio emission in two distinct orbits suggest a pure thermal emission along the orbit, in agreement with the expectation from the extension of the radio photosphere. As a result, we do not see how to reconcile our new measurements with the composite or non-thermal emission reported by \citet{montes2009}, in the context of the binary scenario with the 112.4-d period. 

\subsubsection{The triple system scenario}\label{disctriple}

The catalogue of PACWBs \citep{catapacwb} includes several hierarchical triple systems, with the measured non-thermal radio emission coming from the WCR of the outer, wider orbit. It is worth to investigate how such a scenario may be compatible with the results of radio observations of WR\,133, though no hint for a third object exists to date at other wavebands. In such a scenario, one may distinguish two orbits: the so-called inner orbit as illustrated in Fig.\,\ref{orbit}, and the hypothetical outer orbit. Both are characterized by their own period and eccentricity, and are associated to a WCR. If the size of the inner orbit is too small to warrant a detection of synchrotron emission from its WCR (see the above discussion on the free-free absorption), the wind collision in the outer orbit may be located far enough from the inner parts of the WR wind to be significantly unveiled in a significant part of the orbit. In this scenario, the lack of detection of non-thermal radio emission in 2015 may be explained by a system close enough to the outer periastron passage, when the outer WCR reaches deeper layers of the opaque stellar winds of the inner system. 

\begin{figure}[ht]
\begin{center}
\includegraphics[width=85mm]{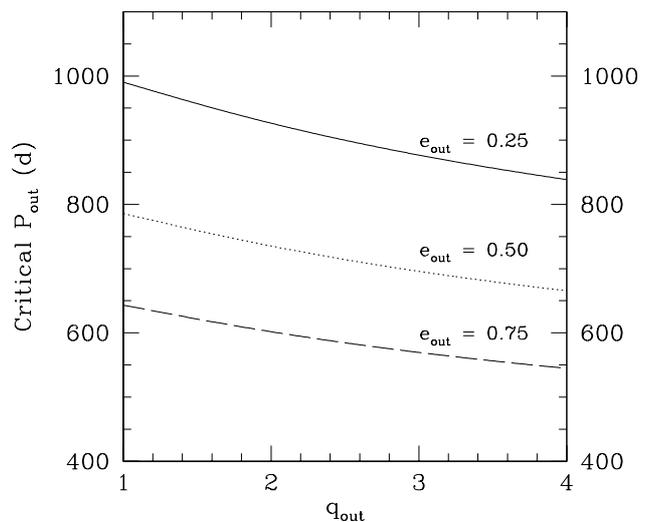}
\caption{Estimate of the critical minimum outer period of WR\,133 allowing for a dynamical stability of a hierarchical triple system. The critical period ($P_{out}$) is plotted as function of the outer mass ratio ($q_{out}$), assuming three values for the outer orbit eccentricity.\label{wr133dyn}}
\end{center}
\end{figure}

The existence of such a hypothetical triple system raises the question of its dynamical stability. To address this question, we made use of the stability criterion proposed by \citet{triplestability}. This criterion expresses the critical, minimal orbital period ratio (outer/inner) allowing for the dynamical stability of a triple system, as a function of the mass ratio of the inner and outer systems, and of their respective eccentricities. Using equations 1 to 3 in \citet{triplestability}, we computed the critical period of the outer orbit for a range of mass ratio of the outer system ($q_{out}$), assuming different arbitrary values for the eccentricity of the outer orbit ($e_{out}$ = 0.25, 0.50 and 0.75). The results are plotted in Fig.\,\ref{wr133dyn}. For $q_{out}$, we considered somewhat conservative values. The minimum boundary is 1, which is quite optimistic. A third object as massive as the WN5 + O9I pair would be bright enough to have been easily revealed by previous spectroscopic observations. For the maximum boundary, a value of 4 would be adequate for instance if the third component is an early B object ($m_3$ $\sim$ 11-12 M$_\odot$). Such a star would still produce a stellar wind able to sustain a significant wind-wind interaction for non-thermal physics, and may be faint enough as compared to the two other components to remain undetected so far. Intermediate values of $q_{out}$ could stand for late-type O-stars, especially on the main-sequence for high values of the mass ratio. One sees that outer periods in such a hypothetical triple system should be of at least several hundreds of days to warrant dynamical stability. Provided the outer orbit is long enough there is a priori no physical reason to consider such a triple system is unlikely.

The outer orbit must be wide enough to warrant a significant detection of non-thermal emission, but in addition it should be significantly eccentric to explain the very different behaviors reported between the epochs of the \citet{montes2009} measurements (i.e., year 1993) and our observations. Close to periastron, the wind-wind interaction region (where electrons are certainly accelerated) is located closer to the inner pair. The synchrotron emission region is thus more deeply embedded into the dense WN5 + O9I winds, leading to a more severe free-free absorption of the synchrotron spectrum. At orbital phases far enough from periastron, the wind-wind interaction will be mainly located outside the radio photosphere and some composite radio emission (thermal + non-thermal) may be detected, as was the case for the 1993 measurements. 

\begin{table*}
\caption{Comparison between the binary and triple system scenarios for WR\,133. \label{comparison}}
\begin{center}
\begin{tabular}{l l l}
\hline
Feature & Binary system scenario & Triple system scenario  \\
\hline
\vspace*{-0.2cm}\\
Detection of non-thermal & Difficult to reconcile with & The larger size of an eccentric \\
emission despite a strong & the measurement of a thermal  &  outer orbit should allow the \\
free-free absorption. &  spectrum across the 112.4-d  & detection of non-thermal emission\\
 & orbit. & at some epochs. \\ 
 \vspace*{-0.2cm}\\
Switch between detection and & Difficult to explain in the & May be explained by an eccentric \\
non-detection of non-thermal & context of the 112.4-d orbit & outer orbit with observations \\
emission with a time interval & considering the lack of detection & closer to periastron passage \\
of several years.  & in our sampling of the orbit. & in July 2014 and September 2015. \\
\vspace*{-0.2cm}\\
Lack of evidence for a & In favor of the binary system & Not in favor of the triple system \\
third object in the system. & scenario. & scenario, but provides loose \\
  &  &  constraints on the nature of  \\
  &  &  the hypothetical third object,\\
  &  &  and previous studies are not \\
  &  &  sufficient to completely reject \\
  &  &  the existence of a third star.\\
\vspace*{-0.2cm}\\
\hline
\end{tabular}
\end{center}
\end{table*}

Finally, let us emphasize that any direct indication of the presence of a third component in the system is still lacking to date. At this stage, the triple system scenario is envisaged for the sake of completeness, and it may be supported or rejected according to future observations of WR\,133. One has however to remind that the late detection of a third component orbiting a previously known massive binary system is not a scarce event (some of these examples are included in the catalogue of PACWBs, \citealt{catapacwb}). For a quite eccentric orbit, the amplitude of the radial velocity variations may be quite low along a large part of the orbit, with measurable changes while approaching periastron passage. A low inclination of the outer orbit would also inhibit the detection of the putative third object. Let's note also that the time sampling of the radial velocity measurements used by \citet{UH1994wr133} to compute the orbital solution is quite sparse, for a time basis of about six years. Such a sampling leaves significant room for a still undetected companion on a wide eccentric orbit. This is especially true considering the difficulty to identify the spectroscopic signature of companions in multiple systems harboring a Wolf-Rayet star. In this context, especially considering the likely long period of the putative third component, long baseline interferometry constitutes a very adequate technique, as demonstrated for instance by \citet{LeBouquin} for a sample O-type binaries. This would allow us to search for an astrometric companion, and to monitor the expected orbital motion over long time scales (i.e., up to several years).

\subsection{On the identification of synchrotron radio emitters}\label{bias}

The switch between a thermal and a composite radio spectrum is especially relevant for the discussion of the detectability of synchrotron radiation in massive colliding-wind binaries. At best, a non-thermal/composite spectrum is identified at one epoch. Any observation of that system when the stellar separation is short enough to bury deeper the synchrotron emission region in the combined stellar winds will only reveal a purely thermal spectrum. In the hypothetical triple system scenario, the orbital phase interval where non-thermal emission should be detectable is wider, but observations reported here show that it is not detected in July 2014 and September 2015. This clearly indicates how critical observational biases affect our capability to detect non-thermal radio emission in these systems, and therefore to attribute a particle accelerator status.

The issue of observational biases is critical for the identification of PACWBs, as discussed by \citet{DBRP}. Several systems already observed with radio observatories and displaying a thermal spectrum may reveal a non-thermal emission component at other orbital phases. On top of that, the inclination of the system and the longitude of ascending node will dictate which stellar wind contributes the most to the line-of-sight absorbing material at any orbital phase. This orientation effect is especially sensitive to dense and opaque winds of Wolf-Rayet stars, along with early supergiant winds with enhanced mass loss. This has two essential consequences. First of all, the quest for new particle accelerators to be included in the PACWB catalogue should proceed through multiepoch campaigns, improving the probability to observe a system at an orbital phase where the synchrotron emission may be detected. For systems with a well-characterized orbit, radio observations may focus on a few selected orbital phases. Second, this observational bias strongly suggests that the fraction of PACWBs among massive binaries is significantly underestimated by the present census.

\subsection{Radio emission from a nearby source}\label{radionearby}
The analysis of the radio images revealed the presence of a point source at about 10 arcsec northeast from WR~133, at coordinates [20:05:57.70; 35:47:26.05]. The closest counterpart in the 2MASS catalogue \citep{2mass2006} is 2MASS\,J20055779+3547287, at about 3 arcsec of the NE source. According to \citet{masonspeckle}, a few visual companions close to WR\,133 are identified in the Washington Double Star catalogue (WDS), at angular distances ranging between 6 and 36 arcsec. None of these visual companions appear at a position angle of 30-40$^\circ$ with a separation of about 10 arcsec \footnote{The WDS catalogue data can be accessed via the following link: http://www.usno.navy.mil/USNO/astrometry/optical-IR-prod/wds/WDS}. 

Following the same procedure as for WR~133, we determined the flux density in both bands and we derived the spectral index (see Table\,\ref{other}). This source seems to be constant and presents a non-thermal spectrum at both epochs, strongly suggesting the emission is significantly made of synchrotron radiation. 

Assuming the source is of Galactic origin, one can only speculate on its nature. One may mention a couple of scenarios which require additional information to be checked consistently. The source could either be a massive binary presenting the same behavior as PACWBs \citep[e.g.,][]{catapacwb}, or a young stellar object presenting non-thermal emission associated to jets interacting with the interstellar medium \citep[e.g.,][]{HH80812017}. Negative spectral indices are also reminiscent of supernova remnants (SNRs) known as emblematic Galactic synchrotron sources \citep[e.g.,][]{dubner2015}. Even though most SNRs are characterized by spectral indices close to $-0.5$, some of them display values up to $-0.3$. Such flatter spectra are sometimes measured for older SNRs, even though the physical justification for such a value is not straightforward \citep{RGB2012}. However, if known SNRs quoted in Green's catalogue were moved to distances such that their flux density (actually, of the order of 1--10\,Jy) would be as low as a fraction of mJy, one would have to push them beyond the limit of the Milky Way. This does not argue in favor of a Galactic SNR for this unidentified source.

\begin{table}
\caption{Flux density measurements of the NE source.\label{other}}
\begin{center}
\begin{tabular}{l c c c}
\hline
Epoch & S$_{\rm 5.5\, GHz}$ & S$_{\rm 9\,GHz}$ & $\alpha$ \\
 & (mJy) & (mJy) & \\
\hline
\vspace*{-0.2cm}\\
27 Feb 2014 & 0.43\,$\pm$\,0.02 & -- & -- \\
28 Feb 2014 & 0.42\,$\pm$\,0.02 & -- & -- \\
13 Apr 2014 & 0.43\,$\pm$\,0.02 & -- & -- \\
23 May 2014 & 0.41\,$\pm$\,0.02 & -- & -- \\
24 Jul 2015 & 0.43\,$\pm$\,0.02 & 0.37\,$\pm$\,0.01 & $-0.31\,\pm$\,0.11 \\
20 Sep 2015 & 0.47\,$\pm$\,0.03 & 0.40\,$\pm$\,0.04 & $-0.33\,\pm$\,0.24 \\
\vspace*{-0.2cm}\\
\hline
\end{tabular}
\end{center}
\end{table}

\section{Summary and conclusions}\label{concl}
We investigated the radio emission from the WN + O binary WR\,133 on the basis of dedicated JVLA observations at 5.5 and 9 GHz, at various epochs spread over two different orbits. Our measurements reveal a thermal spectrum, in apparent contradiction with the detection of a non-thermal/composite spectrum reported by \citet{montes2009}. 

In the context of a simple binary system scenario, our monitoring of the 112.4-d orbit rejects the idea that a non-thermal emission component could be detected. This can easily be explained by the strong free-free absorption due to the dense stellar winds, constituting a severe obstacle for the emergence of a measurable non-thermal radio contribution from the wind collision region. However, this does not provide any satisfactory interpretation context for the non-thermal emission reported previously. Alternatively, one may think about a triple hierarchical system scenario. In this interpretation framework, the hypothetical third object would move along on a wide outer orbit. The outer wind collision region would be much less affected by the free-free absorption, opening the possibility that a non-thermal emission component may be detectable at some epochs provided the outer orbit is sufficiently eccentric. According to simple dynamical stability considerations, the outer orbital period should be at least several hundred days. The main drawback of the triple system interpretation is the present lack of any hint for a third object in this system, though (i) previous investigations of other systems illustrated the difficulty to identify high multiplicity systems, and (ii) previous spectroscopic studies of WR\,133 are not sufficient to completely reject the triple system scenario. The search for a putative third object on a wider orbit would require either dedicated spectroscopic measurements over a long time basis, or more likely visible/infrared interferometric observations using long baseline interferometers to search for an astrometric companion.  

Beside the question of the multiplicity (binary or triple) of WR\,133, the change in the spectral shape (with a composite spectrum at one epoch) provides evidence for the existence of a synchrotron source, despite the strong free-free absorption at work in this system. The epoch-dependent detection of the non-thermal emission constitutes a compelling illustration of the observational biases that severely affect the detectability of synchrotron sources in colliding-wind massive binaries. The depth of the location of the synchrotron emission region in the combined stellar winds and orientation effects can represent significant obstacles for the detection. A system may present the properties of a purely thermal emitter in a significant part of its orbit, with only hints for synchrotron emission in another orbital phase interval. One can thus anticipate that the actual population of particle accelerators among colliding-wind binaries is certainly underestimated. Observation strategies aimed at unveiling new particle accelerators should take this issue into account notably through an adequate selection of orbital phases (when feasible), or at least by organizing multiepoch observations to sample the orbit.

Finally, our JVLA observations revealed the existence of non-thermal point source, about 10 arcsec northeast from the position of WR\,133. The source seems constant, as far as our time sampling allows to say. The nature of the object is still undetermined and it may be related to a Galactic stellar source such as another colliding-wind binary or a young stellar object.

\begin{acknowledgements}
The authors want to express their gratitude to the NRAO staff for the scheduling of the VLA observations, to Drew Medlin, Rick Perley and the NRAO helpdesk. Our acknowledgements go also to Dr. Juan-Ramon Sanchez-Sutil, for his previous help in the preparing of the telescope time proposals and observations and to C. Hales for fruitful discussions on data reduction techniques. The authors also thank the anonymous referee for his/her constructive report that allows to significantly strengthen our conclusions. The National Radio Astronomy Observatory is a facility of the National Science Foundation operated under cooperative agreement by Associated Universities, Inc. MDB acknowledges the financial support from a ULi\`ege 'Cr\'edit classique (DARA)'. The SIMBAD database was used for the bibliography. This research has made use of the Washington Double Star Catalog maintained at the U.S. Naval Observatory.
\end{acknowledgements}

\bibliographystyle{aa}



\end{document}